
\magnification=\magstep1
\baselineskip=16pt
\vsize=8.9 true in
\hsize=6.5 true in
\hoffset=-0.1truecm\voffset=0.3truecm
\nopagenumbers\parindent=0pt
\footline={\ifnum\pageno<1 \hss\thinspace\hss
    \else\hss\folio\hss \fi}
\pageno=-1

\newdimen\windowhsize \windowhsize=13.1truecm
\newdimen\windowvsize \windowvsize=6.6truecm

\def\heading#1{
    \vskip0pt plus6\baselineskip\penalty-250\vskip0pt plus-6\baselineskip
    \vskip2\baselineskip\vskip 0pt plus 3pt minus 3pt
    \centerline{\bf#1}
    \global\count11=0\nobreak\vskip\baselineskip}
\count10=0
\def\section#1{
    \vskip0pt plus6\baselineskip\penalty-250\vskip0pt plus-6\baselineskip
    \vskip2\baselineskip plus 3pt minus 3pt
    \global\advance\count10 by 1
    \centerline{\expandafter{\number\count10}.\ \bf{#1}}
    \global\count11=0\nobreak\vskip\baselineskip}
\def\subsection#1{
    \vskip0pt plus3\baselineskip\penalty-200\vskip0pt plus-3\baselineskip
    \vskip1\baselineskip plus 3pt minus 3pt
    \global\advance\count11 by 1
    \centerline{{\it {\number\count10}.{\number\count11}\/})\ \it #1}}
\def\firstsubsection#1{
    \vskip0pt plus3\baselineskip\penalty-200\vskip0pt plus-3\baselineskip
    \vskip 0pt plus 3pt minus 3pt
    \global\advance\count11 by 1
    \centerline{{\it {\number\count10}.{\number\count11}\/})\ \it #1}}

\def\eol{\hfil\break}
\def\affl#1{\noindent\llap{$^{#1}$}}
\def\simlt{\lower.5ex\hbox{$\; \buildrel < \over \sim \;$}}
\def\simgt{\lower.5ex\hbox{$\; \buildrel > \over \sim \;$}}

{
\def\cl#1{\hbox to \windowhsize{\hfill#1\hfill}}
\hbox to\hsize{\hfill\hbox{\vbox to\windowvsize{\vfill
\bf
\cl{HI SPIN TEMPERATURES AND HEATING}
\cl{REQUIREMENTS IN OUTER REGIONS OF}
\cl{DISK GALAXIES}
\bigskip
\cl{Edvige~Corbelli$^{1}$ and Edwin~E.~Salpeter$^{2}$}
\bigskip\rm
\cl{Preprint n.~24/93}

\vfill}}\hfill}}

\vskip5truecm
{\leftskip1.7truecm
\affl{1}Osservatorio Astrofisico di Arcetri,
\eol
Largo E.~Fermi 5, I-50125 Firenze (Italy)
\bigskip
\affl{2}Center for Radiophysics and Space Research,
\eol
Cornell University, Ithaca NY, 14853 (USA)

\vfill
To appear in The Astrophysical Journal (Dec. 10, 1993).
\vglue3truecm
}
\eject

\vglue 5 true cm
\heading{ABSTRACT}

\vglue 0.2 true in

We show how to use 21-cm
emission and absorption studies to estimate the heat inputs to the
neutral gas in low pressure environments, such as in outer
disks of spiral galaxies, in galactic halos or in
intergalactic space.
For a range of model parameters we calculate the gas kinetic
and spin temperatures ($T_K$ and $T_S$) and the relation between
$T_S$ and the heat input to the gas. We outline the conditions
for a ``two phase medium'' to exist.
We find that although $T_S$ can be much smaller than $T_K$,
$T_S$ is always $ \gg 3$ K for column densities greater that
$5 \times 10^{18}$ cm$^{-2}$.
This excludes the possibility that relevant HI masses at the
periphery of galaxies are invisible at 21-cm in emission.
Therefore sharp HI edges, observed in outer
disks near column densities $N_l \sim 2\times 10^{19}$ cm$^{-2}$,
cannot be ``fictitious'' edges due to a sudden
decrease of the 21-cm brightness.
The outermost interstellar gas in a disk galaxy is more directly
affected by external processes and in this paper we estimate the
intensity of the extragalactic background at energies close
to 0.1 keV by comparing our theoretical results with HI
emission/absorption studies.  We take into account the
possibility that some energy produced in the inner regions
affects the energy balance in outer regions.
We find that in the absence of any other local heat source
QSO dominated background models are still compatible with
the spin temperature limits derived for
the two best documented HI emission/absorption studies in
outer regions.
However, if future observations should establish that the spin
temperature
is as high as 1000 K, then relevant energy inputs from local
sources become necessary.

\medskip
\noindent\underbar{\strut Subject}\ \underbar{\strut headings}:
Cosmic Background Radiation -- Galaxies:ISM --
Radio Sources: 21 cm Radiation.

\vglue\windowvsize plus 1fill minus \vsize
\eject

\pageno=1
\section{ INTRODUCTION}

\vglue 0.1 true in

Neutral hydrogen exists outside the stellar region of spiral galaxies
as extended disks, coplanar or tilted with respect to the innermost
luminous disk. Isolated concentrations, clouds or
plumes have sometimes been detected well outside the HI disk
(see for example Giovanelli $\&$ Haynes 1988 for a review).
These blobs may indicate a recent disturbance in the HI
distribution due to tidal interactions with a companion or to a strong
burst of activity in the inner disk. In our Galaxy smaller blobs of
HI, closer to the luminous disk, and known as high velocity clouds
(hereafter HVC), have also been detected (Giovanelli 1980).

The physical conditions of the gas in the outer regions of
spiral galaxies are not constrained as they are
in the interstellar medium. In
the inner disk of our Galaxy, where star formation takes place,
the following conditions hold:
{\it (i)}neutral hydrogen column densities are of
order $N_{HI} \sim 2 \times 10^{20} - 10^{21} $cm$^{-2}$, and
extragalactic UV or soft X-ray photons with energies between 13.6 and
200 eV hardly penetrate this layer;
{\it (ii)}even without magnetic pressure and bulk motion ram pressure,
the gas thermal pressure is fairly large, $P/k \sim 3000$ cm
$^{-3}$ K ($k$ is the Boltzmann constant);
{\it (iii)}neutral hydrogen
exists in a cold phase at the kinetic temperature $T_K\sim 80$
and in a warm phase
at $ T_K \sim 8,000$ K which fills a non trivial part of our ISM
($\ge 30\%$ Brinks 1990; Knapp 1990).The ionization of the gas in
the midplane might be due to O and B starlight (Kulkarni $\&$
Heiles 1988), and the
mechanical energy input to the gas from O - star winds and
from blastwaves produced by supernovae (McKee $\&$ Ostriker 1977).
However these hot stars cannot easily account for the
diffuse warm component observed at a few hundreds parsecs
above the galactic midplane (Reynolds 1990).

In outer disks, or in high velocity clouds, stars are absent and the
ionizing flux as well as the mechanical energy input, from the
stellar disk of the galaxy might be very small.
If chains of supernovae feed galactic fountains which reach
the halo directly (Heiles 1987; Corbelli $\&$  Salpeter 1988;
MacLow $\&$ McCray 1988; Martin $\&$ Bowyer 1990) or if there
is hydromagnetic wave dissipation
(Ferriere, Zweibel $\&$ Shull 1988) then the non radiative energy
input into the outer regions are important.
Ionizing flux from extragalactic sources is likely to be present
due to thermal free-free emission from hot gas, to quasars, or to
more exotic sources like dark matter decay (Melott, McKay, $\&$
Ralston 1988; Sciama 1990) or hidden X-UV emitters.
The far UV and soft X-ray emission from quasars is of interest
here because we mainly consider regions with $N_{HI} \le 5
\times 10^{19}$ cm$^{-2}$  and
photons below 0.2 keV, which do not affect the inner disk,
penetrate these smaller column densities and give a large ionization
rate. We do not discuss which ionizing fluxes and
non radiative
heat inputs are more likely on physical grounds but we parametrize
the overall ionizing flux and the non radiative energy input separately
and calculate their effects on the HI gas.
Using HI observations we can then investigate indirectly the
energetic environment of spiral galaxies.
The absence of stars in outer regions makes
self-gravity of the gas layer an important force which
has to be considered, together with dark matter and coronal gas,
in equilibrium models.
Nevertheless we expect a much lower thermal pressure
than that measured in the inner disk, as suggested by
the flaring of the outer disk observed in external galaxies and in
our own Galaxy (Merrifield 1992).

HI absorption studies both on outer disks and on HVC,
show that only a very small fraction of the gas has low
enough temperature to be detected in absorption (Corbelli $\&$
Schneider 1990, hereafter CS; Colgan, Salpeter, $\&$ Terzian 1990,
hereafter
CST; Wakker, Vijfschaft, $\&$ Schwarz 1991; Carilli, van Gorkom,
$\&$ Stoke 1989; Carilli $\&$ and van Gorkom 1992; Schneider $\&$
Corbelli 1993).
However sensitive searches of HI emission at some locations where
no absorption measurements are possible,
indicate that there is a transitional HI column density $N_l$ such
that appreciable emission is seen for $N_B \simgt N_l$ but little
emission is seen for $N_B < N_l$. We use the symbol $N_B$
to indicate the column density
derived directly from the 21-cm brightness temperature;
$N_B = N_{HI}$ when $T_S$ is well above the
background radiation temperature, $T_R$, otherwise
$N_B < N_{HI}$. The transitional column density $N_l$
is of the order of $5 \times
10^{18}$ cm$^{-2}$ for high velocity clouds and is of the
order of $2 \times 10^{19}$ cm$^{-2}$ for outer disks
(CST; Corbelli, Schneider, $\&$ Salpeter 1989) or slightly
smaller (Hoffman {\it et al.} 1993).
In this paper we discuss (and then eliminate) the possibility
that, due to
low pressure conditions and to low photoionization rate in outer
regions, when $N_{HI}$ gets below $N_l$,
the spin temperature strongly deviates from the gas kinetic
temperature and, while $T_S$ approaches $T_R$,
$N_B$ is depressed with respect to $N_{HI}$.
Fictitious HI edges would then appear if one assumes
that $N_B \simeq
N_{HI}$. The lack of HI absorption lines in outer regions
already suggests that ``subthermal effects'' cannot hide
appreciable HI masses, but we shall show more quantitatively
that subthermal effects  are not responsible for
HI edge appearances. A totally different possibility, and a
likely explanation for HI edges, is that they are sharp HI-HII
transitions due to some ionizing flux. This will be analyzed
in detail by Corbelli and Salpeter (1993, hereafter
Paper II). Nevertheless due to the uncertainties in the evaluation
of HI masses in extragalactic objects where no absorption measures
are possible (Giovanelli $\&$ Haynes 1991; Schneider {\it et al.}
1989) and also to the importance of subthermal effects
for HI absorption lines, we shall present here a full
calculation of the spin temperature for several heating
inputs outside the galaxy.

In Section 2 we introduce some models for
the extragalactic soft X-ray background radiation
and in Section 3 we display the equilibrium equations
and parameters used for modeling the gas distribution outside the
stellar disk. Section 4 shows results relative to
subthermal effects in the 21-cm emission line
and in Section 5 we derive in more generality
the relation between the kinetic and spin temperature
as a function of the background radiation intensity,
and of additional non ionizing local heat sources.
We outline the conditions of the HI in a low density
medium pervaded with decaying neutrinos
in Section 6. The last Section summarizes  constraints
on the energetic environment of spiral galaxies which can be
set by comparing our theoretical results with 21 cm observations.

\vglue 0.1 true in

\section
{LOCAL ENERGY INPUTS AND THE EXTRAGALACTIC SOFT X-RAY}
{\bf BACKGROUND}

\vglue 0.4 true in

As discussed in the introduction, there may be energy inputs
which are non radiative and non ionizing and which
may vary from one galaxy to another or from one region to
another (tidal
interaction, hydromagnetic wave heating, fountains, etc.).
We shall characterize these inputs using a single number,
$E_{nr}$, the heat per hydrogen atom per second due to non
photoionizing sources in units of eV s$^{-1}$, which should
be added to the heat input from ionizing radiation, $E_r$.
The ionization-recombination equilibrium depends on the photon
spectrum $N(E)$.
Below we summarize the known upper or lower limits to the
extragalactic background radiation at zero redshift
and then we will parameterize various model assumptions.

The UV and soft X-ray background has been reviewed by Bowyer
(1991) and by McCammon, $\&$ Sanders (1990). Above $\sim$ 2 keV
a uniform extragalactic X-ray flux of power law
form is known to exist (Schwartz 1978):

$$ {{\hbox {d}}N \over {\hbox {d}}E} = 7.7 E^{-1.4}_{keV}
\qquad {\hbox {ph cm$^{-2}$ s$^{-1}$ sr$^{-1}$ keV$^{-1}$ }}
\eqno (1)$$

Between 2 keV and 0.5 keV the total X-ray background is about
3$\times 10^{-8}$ erg cm$^{-2}$ s$^{-1}$ sr$^{-1}$
(Shanks {\it et al.} 1991; Wu {\it et al.} 1991),
higher than an extrapolation of eq. (1) to this
energy range, but the extragalactic contribution
is uncertain because of local emission,
and it is not measurable directly at all
below 0.3 keV because of absorption by the local HI disk.
The space density and luminosity function for quasars
are known for redshifts $z<2$ (see e.g. Hartwick and Schade
1990) and their contribution to the flux between 2 keV and 0.5 keV
has been estimated.
Preliminary analysis of ROSAT deep survey images shows that
quasars can account for at least 20$\%$ of the measured X-ray
background below 2 keV (Hasinger, Schmidt,
$\&$ Trumper 1991; Wu, $\&$ Anderson 1992)
and therefore we consider 5$\times 10^{-9}$ erg cm$^{-2}$
s$^{-1}$ sr$^{-1}$  as a definite lower limit for the extragalactic
background intensity between 2 kev and 0.5 keV.
An extrapolation of eq. (1) down to 1.5 keV yields this
minimum expected intensity which we consider in examining
the possibility of strong subthermal effects in emission.

In this paper we use a few spectral models of the
extragalactic background radiation down to 0.1 keV.
If this background is dominated by quasars, its spectral shape
in the absence of attenuation is approximately of the form:

$$ {{\hbox {d}}N \over {\hbox {d}}E} = I\times E^{-2.4}_{\hbox{keV}}
\qquad {\hbox {ph cm$^{-2}$ s$^{-1}$ sr$^{-1}$ keV$^{-1}$ }}
\eqno (2)$$

\noindent
which we shall refer to as the ``quasar spectrum''.
For $I\simeq 15$  the
spectrum described by eq. (2) corresponds to a flux density
$\simeq 4\times 10^{-23}(E_{\hbox{eV}}/13.6)^{-1.4}$
erg cm$^{-2}$ s$^{-1}$ sr$^{-1}$ Hz $^{-1}$. This flux at
zero redshift
was suggested by Sargent {\it et al}. (1979) and connects the
so called ``blue bump'', observed in QSO spectra, with the cosmic
background flux observed above 2 keV. In a recent
paper Madau (1992) estimates a lower value of the integrated
UV cosmic background from observed QSOs. He takes into account
the possible attenuation of the flux due to intervening
Lyman-$\alpha$ clouds. However above 0.1 keV attenuation is
probably unimportant and Madau's estimate of the unattenuated flux
down to a cutoff energy of 0.1 keV is given by
the quasar spectrum of equation (2) with $I\simeq 5$.

We shall consider three different cases for flux and energy inputs.
For Case A we assume the minimum extragalactic X-ray flux in order
to compute the most extreme subthermal effects.
We use a more likely flux for Case B and
Case C and show how to use HI spin temperature measurements
to constrain the extragalactic background flux or the local
heating just outside disk galaxies.

$\underline {\hbox{Case A}}$:
We set $E_{nr}$=0 and use the flux described by
equation (1) with a lower cut-off energy $E_c=1.5$ keV or
$E_c=0.5$ keV (in this last case
about 60$\%$ of the observed background flux between 2 and 0.5
keV is of extragalactic origin).

$\underline {\hbox{Case B}}$: We consider  $E_{nr} \ne 0$
as an adjustable parameter and use a low
quasar flux, namely equation (1) with a cut off energy $E_c=0.1$
keV.

$\underline {\hbox{Case C}}$: We add a soft X-ray component in
the energy range $E_c<E<1.5$ keV, to
the cosmic spectrum given by eq. (1).
The following form describes the spectral law which we
use in this case :

$$ {{\hbox {d}}N \over {\hbox {d}}E} =
\left\{ \matrix {7.7 \times E_{keV}^{-1.4}\qquad E\ge\ 1.5\ {\hbox {keV}}
\hfill\cr  A\ E_{keV}^{-b} \qquad 0.5\le E<1.5 {\hbox {keV}}
\hfill\cr  I\ E_{keV}^{-s}
\qquad E_c\le E < 0.5\ {\hbox {keV}} \hfill\cr
0\ \qquad \qquad E < E_c \hfill\cr}\right\}
{\hbox{ph}}\ {\hbox{cm}}^{-2} {\hbox{s}}^{-1}
{\hbox{sr}}^{-1} {\hbox{keV}}^{-1}\eqno(3)$$

\noindent
where $A$ and $b$ are constants determined by requiring d$N$/d$E$
to be continuous at 0.5 and at 1.5 keV and $I$ is an adjustable
parameter. In this paper we use $E_c=0.1$ keV,
$s=2.4$, and $E_{nr}=0$ unless stated otherwise.
The quasar spectrum in eq. (2) can be represented to
a good approximation by eq. (3) using $s=2.4$ (and $I\sim 5$
for Madau's intensity).

In Fig. 1 we show d$N$/d$E$ for some spectra used.
The background spectrum at energies close to the Lyman limit
will not be of interest in this paper because here we
consider the heating of slabs with HI column densities
$\simgt 0.5 \times 10^{19}$ cm$^{-2}$. However
the complete absence of photons below 0.1 keV,
is an oversimplification; some photons
below 0.1 keV will penetrate the gas with low HI column
density and increase the ionization rate slightly.
In Section 6 we discuss $T_S$ and $T_K$ for
the case of monochromatic photons with $E\sim14.5$ eV
which are produced locally throughout the disk.

\vglue 0.1 true in

\section
{CHARACTERISTIC PARAMETERS OF HI GAS OUTSIDE A GALAXY}

\vglue 0.1 in

The brightness temperature of the 21-cm emission
line, $T_B$, is related to the  column density of neutral
hydrogen along the line of sight, $N_{HI}$:

$$T_B \propto N_B \equiv \chi_B N_{HI},\qquad \chi_B \equiv
{T_S - T_R\over T_S}   \eqno (4)$$

\noindent
where $T_R=2.73 $ K is the temperature of the microwave background
radiation, $T_S$ is the spin temperature, defined through
the population of the two hyperfine levels of neutral hydrogen
(Field 1958).
The factor $\chi_B$ is close to unity if the kinetic
temperature, $T_K$ is well above the cosmic background temperature,
and the medium is in LTE ($T_S \simeq T_K$). In this case the
brightness temperature is directly proportional to $N_{HI}$.
The spin temperature is important
for the optical depth at the line center $\tau_c$
which can be determined through absorption measurements

$$ \tau_c = {5.14 \times 10^{19} \over \omega} {N_{HI}
\over T_S} \eqno (5)$$

\noindent
where $\omega$ is the width of the line at half maximum in
km s$^{-1}$. Most absorption and emission studies give $T_S$
and $N_{HI}$ and not the kinetic temperature of the medium
because $\omega$ might not be proportional to $\sqrt{T_K}$, like
if there is turbulence in the medium or the absorption line
is unresolved. If $\omega$ is unknown
we use the following value:

$$\omega \simeq {\hbox {max}} (4, 0.21\sqrt{T_K})\ {\hbox {km s}}^{-1}
\eqno (6)$$

\noindent
based on the line width of the few detected absorption lines
(Rubin, Thonnard, $\&$ Ford 1982; Carilli $\&$ van Gorkom 1992).
Deviations of $T_S$ from $T_K$ are especially important
in outer regions where low densities slow down the
collisional excitation of hyperfine levels of H.
We calculate $T_S$ for all cases in
this paper using equation (15) of Field (1958).
We consider Lyman-$\alpha$ pumping through
photons generated by $H^+ -e$ recombinations (Spitzer 1978),
by bound-bound transitions due to thermalized electrons
(Spitzer 1978), and  by bound-bound transitions due
to non thermalized electrons coming from H photoionization
(Shull $\&$ Van Steenberg 1985).
Following Deguchi $\&$ Watson (1985) we set
the Lyman-$\alpha$ temperature $T_L$ equal to $T_K$
and use the expression  of Bonilha {\it el al.} (1979) for
the number of Lyman-$\alpha$ photon scatters inside the
cloud, disregarding the possibility of strong velocity gradients.
For the collisional de-excitation probability of the hyperfine
levels via neutral and ionized H atoms, and via electrons, we
use the expression given by Smith (1966).
For $T_R << T_S << T_K$ the gas is still far from thermodynamic
equilibrium but the factor $\chi_B$ is very close to unity.
This means that
subthermal effects show up much more easily in
absorption rather than in emission measures, because
emission depends on the ratio between the population of the
two hyperfine levels which remains $\simeq 3$.
Absorption depends on their population difference which can
vary by a much larger factor.

The gas will be considered distributed in a slab, with the
vertical extension much smaller than the horizontal one.
This is a good representation for extended HI disks or flat clouds
in the outer parts of galaxies. For more spherical blobs our
calculation will be indicative but not rigorously correct.
The vertical distribution of gas density depends on the ratio of
dark matter to gas and on any contribution of magnetic fields to
pressure (or to buoyancy). These effects will be discussed in
detail in Paper II; in this paper we assume that the
ideal gas law connects pressure, $P$, to gas
kinetic temperature, $T_K$. Furthermore we shall use here the
single slab approximation.
For fixed values of the HI column density we compute the
kinetic temperature and subthermal effects at a height $z=z_{1/2}$
above the central plane such that half the gas mass per unit area
lies between $-z_{1/2}$ and $+z_{1/2}$. The attenuation of
the extragalactic flux due to photoionization of HI, HeI, and
HeII, is computed assuming that the fractional ionization of
HeI and HeII at all levels is the same as
at $z_{1/2}$.

The real pressure will have an additional term due to
the external pressure, $P_{ext}$, and to the compression
from a dark matter distribution.
Charlton, Salpeter $\&$ Hogan (1993) estimate that the purely
extragalactic value for $P_{ext}/k$ is less than $\sim 4$
cm$^{-3}$ K but some galactic coronal gas is likely
to increase the effective value of $P_{ext}$.
For a spherical halo which gives a rotational velocity
of 100 km/s at $R=15$ kpc, the dark matter compression is
equivalent to about 70 cm$^{-3}$ K. Since in this
paper we use the uniform slab approximation we mimic the
effects of $P_{ext}$, dark matter and eventually of
some internal non-thermal pressure in one single term,
$P_0$. We use the ideal gas law and we consider
$0.1 \le P_0/k \le 500$ cm$^{-3}$ K. The following
expression describes the total pressure we use for
a given column density of gas (see Spitzer 1942,
Iba\~nez and di Sigalotti 1984):

$$ {P\over k} = {P_0\over k} + {P_{sg}\over k} \simeq
{P_0\over k} + {\pi G m_H^2\over 2.3 k} N_{tot}^2 (1+4h_{He})^2
\simeq {P_0\over k} +
0.36 \Bigl({N_{tot}\over 10^{19}}\Bigr)^2 {\hbox {cm}}^{-3}
{\hbox {K}} \eqno (7) $$

\noindent
where $G$ is the gravitational constant,
$m_H$ is the hydrogen mass, $h_{He}\simeq 0.1$ is
the helium abundance relative to hydrogen, and
$N_{tot}$ is the total (HI+HII) hydrogen column density
(i.e. twice the gas surface density above the equatorial
plane).

We omit dust grains entirely and we
assume that all elements heavier than He have only
half the solar abundance (unless specified differently)
and that there is enough UV light below 13.6 eV,
from the background or scattered outside the stellar disk,
to keep most of the carbon, silicon and iron singly ionized.
For the range of pressures that we consider this is a good
assumption (Boyer 1991).
Then the heating rate due to photoionization
of CI  does not depend on the UV photon intensity but only
on the mean photon energy and we use the value $E_C=2$eV
for the excess energy.
In the temperature range that we consider ($T_R < T_K < 17,000$
K) cooling is dominated by collisional excitation of
CII, SiII, FeII by electrons and HI atoms at low
temperatures, and by excitation of the n=2 level of HI at higher
temperatures. However cooling
by free-free emission and by HII recombinations are
considered as well. $P_0/k \ge 0.1 $ ensures the recombination
time for HII to be shorter than $10^{10}$ years.

For the ionization - recombination equilibrium of HI, HeI, and HeII
we use the ``on the spot approximation'' with the
photoionization cross section of HI, HeI and HeII
as given by Spitzer (1978) and Brown (1971).
We assume that no photons escape from hydrogen recombinations
to the n=1 level (nor from helium recombinations to n$\ge$1)
because they are absorbed elsewhere by other HI atoms. Also, due to
low volume densities in outer regions  we can assume to a good
approximation that all the HeI and HeII recombination photons are
absorbed by HI atoms (Osterbrock 1989).
We include secondary electrons, the heat, and the Lyman
alpha photons produced by primary electrons created via
photoionization of HI, HeI and HeII by using results
of a Montecarlo simulation (Shull $\&$ Van Steenberg 1985).
No secondary electrons are considered when primary electrons
are created via collisions with other atoms. This and the
omission of collisional ionization of helium atoms are good
assumptions since we consider only temperatures
below 17,000 K.
Unfortunately the complete set of results by Shull $\&$
Van Steenberg (1985) is given as function of the
fractional ionization of H only for X-ray energies well
above 0.1 keV. For lower energies they show the fractional
energy deposited as heat and the number of secondary electrons
ejected from hydrogen only for a discrete number of cases.
We use a subroutine which interpolates their results and computes
also the number of secondary electrons ejected from helium atoms,
$\Phi_{He}$, assuming that
$\Phi_{He}/\Phi_H$ is the same as at higher energies. A similar
assumption is used for the fractional
energy of the primary which goes into
Lyman-$\alpha$ excitation for hydrogen (which will be relevant in
computing subthermal effects).
We introduce the following symbols:

\noindent
$f_1(T_K),f_2(T_K)$ are cooling efficiencies due to impact
of neutral hydrogen ($f_1$) and electrons ($f_2$) with heavier
elements (the contribution of each element has been weighted by its
abundance with respect to H).

\noindent
$f_3(T_K)$ is the cooling efficiency for free-free emission
and HI recombinations and $f_4(T_K)$ is
the cooling efficiency due to HI impact with free electrons.

\noindent
$x_H$, $x_{He}$, and $x_{He2}$ are the fractional ionization of HI,
HeI and HeII respectively.

\noindent
$h_{He}=0.1$, $h_C=1.9\times 10^{-4}$ are the helium and carbon
abundance by number with respect to hydrogen.

\noindent
$\alpha_2$ is the HI recombination coefficient excluding
captures to the n=1 level (Spitzer 1978).

\noindent
$\alpha_{He}$ $\alpha_{He2}$ are the HeI and HeII recombination
coefficient (Spitzer 1978).

\noindent
$\alpha_C$ is the total recombination coefficient for CI
which we assume to be equal to that of HI
and $E_C=2$eV is the heat released for each carbon ionization.

\noindent
$\gamma_H$ is the HI ionization rate due to collisions
with free electrons.

\noindent
$\xi_{a1}$, $\Phi_{a1,a2}$ and $E_{a1}$ are respectively:
the numbers of photoionizations per second
per atom of type $a1$, the number of secondary electrons
created by interaction of type $a2$ atoms with primary
electrons coming from photoionization of type $a1$ atoms, and the heat
released for each photoionization of type $a1$ atoms. The
symbol $a1$ can be H, He, or He2 to indicate HI, HeI and HeII atoms,
the symbol $a2$ can be only H or He because secondary electrons
from HeII atoms are not considered.
We compute these quantities at $z_{1/2}$ for isotropic background
radiations described in Section 2,
attenuated by the overlaying and underlaying layers of gas.

\noindent
$n_H\simeq
P/\lbrace k T_K (1+x_H+h_{He}+h_{He}x_{He}+h_{He}x_{He2}+h_C)
\rbrace$ is the volume density of H atoms.

\noindent
$n_e\simeq n_H (x_H+
h_{He}x_{He}+h_{He}x_{He2}+h_C)$ is the volume density of free
electrons.

The ionization equilibrium equations for H and He are then:

$$ n_e \Bigl( {x_H\over 1-x_H} \alpha_2 - \gamma_H \Bigr)
= \xi_H
(1+\Phi_{H,H}+\Phi_{H,He})+ {1-x_{He}-x_{He2}\over 1-x_H}
h_{He}\xi_{He}$$
$$\times (1+\Phi_{He,H}+\Phi_{He,He})+ {x_{He}\over 1-x_H}
h_{He}\xi_{He2} (1+\Phi_{He2,H}+\Phi_{He2,He})\equiv \tilde\zeta
\eqno (8)$$

$$n_e x_{He} \alpha_{He}= (1-x_{He}-x_{He2})
\xi_{He}(1+\Phi_{He,He})+{(1-x_{H})\over h_{He}}\xi_H
\Phi_{H,He}$$
$$ + x_{He}\xi_{He2}\Phi_{He2,He}\eqno(9)$$

$$ n_e x_{He2} \alpha_{He2}  = x_{He2}
\xi_{He2}\eqno (10)$$

\noindent
The right hand side of equation (8) gives $\tilde\zeta$,
the total number
of ionizations per second for a neutral H atom.
If the slab is optically thin to the extragalactic background
eq. (8) gives $\zeta$ the
total number of ionizations per second for an unshielded
neutral H atom. For Case A with a
lower cut-off at 1.5 keV $\zeta\simeq 1.5\times 10^{-20}$
H ionizations s$^{-1}$, while $\zeta \sim 3\times
10^{-17}$ H ionizations s$^{-1}$ for Case C with $E_c=0.1$ keV,
$s=2.4$ and $I\simeq 5$. The ionization rate $\zeta$ for Case
B is about the same as for Case C with $s=2.4$ and $I\approx
1.2$. In paper II we shall extend the photon spectrum down to
the Lyman edge, 13.6 eV, which results in much larger values for
$\zeta$.

The energy equation below describes the balance between
the cooling rate and the heating rate per H atom due to
photoionizations by extragalactic background photons or
to additional local heat sources ($E_{nr}$):

$$\eqalignno {n_H \Big\lbrace (1-x_H) f_1(T_K) &+
\lbrack f_2(T_K) + x_H f_3(T_K) +(1-x_H) f_4(T_K) \rbrack
{n_e\over n_H} \Big\rbrace\cr
=&(1- x_H) E_H \xi_H
+ h_{He}(1- x_{He}-x_{He2})E_{He} \xi_{He}\cr
&+ h_{He} x_{He2} E_{He2}
\xi_{He2} + h_C E_C \alpha_C(T_K)n_e + E_{nr}&(11)\cr}$$

\vglue 0.1 true in

\noindent
The left hand side is the cooling per H atom
minus the heat due to carbon ionization. The right hand side is
the heating per H atom due to photoionization of hydrogen and
helium plus the non radiative heating per atom, $E_{nr}$.
To simplify the solution of the above set of equations
we set $x_{He2}=0$ at low temperatures while at
high temperatures we omit the cooling function $f_1$.
Given $N_{HI}$, $P_0$, and the spectral index of
the background radiation we solve the above equations by expressing
$x_{He}$, $x_{He2}$ and $P/\xi$ as functions of $x_H$ using three
of the four equations given above. For each value of $T_K$ then we
need to solve numerically one single equation of type $f(x_H)=0$.

\noindent
-Case A: due to the high energies of the incoming photons,
and to the low intensity of the radiation (which for this case
is fixed) the gas is optically thin with negligible fractional
ionization of HeII.  Solutions of $f(x_H)=0$ are found assuming
constant values of $\Phi_{a1,a2}$, $E_{a1}$ and $P$. Since
the degree of H ionization is low, $P$ is determined only
by $P_0$ and $N_{HI}$. We calculate $\Phi_{a1,a2}$ and $E_{a1}$
(which are functions of $x_H$) with an iterative procedure.
$T_K$ is determined by the value of $P/\xi$ if the heating rate
due to H and He photoionization is stronger than that due to
carbon photoionization.

\noindent
-Case B: for each value of $T_K$ we guess a value of $E_{nr}$ and
solve iteratively $f(x_H)=0$ to find self consistent values of
$x_H$, $\Phi_{a1,a2}$, $E_{a1}$, and $P$. The
attenuated photon flux (or $\xi$) required at $z_{1/2}$ can be
calculated.
But since for this case the incident photon flux is fixed,
we know what $\xi$ should be for a given $x_H$. We adjust
$E_{nr}$ accordingly and repeat the calculation until we get
the correct value of $\xi$.

\noindent
-Case C: the value of $x_H$, found by solving $f(x_H)=0$,
determines $P/\xi$ and therefore $I$. The correct solution is
found after a few iterations because the spectral distribution
of the extragalactic radiation changes when the parameter $I$ varies
with $T_K$.

For each case we then determine the spin temperature, as described
at the beginning of this Section.

\vglue 0.1 true in

\section
{SUBTHERMAL EFFECTS IN THE 21-cm EMISSION LINE}

\vglue 0.1 true in

If the ionizing flux and the pressure $P$
are moderate or small, the spin temperature $T_S$ is
depressed below the gas kinetic value $T_K$ and this
subthermal effect increases the 21-cm absorption (see Section
5). If flux and pressure were extremely small, $T_S$ could be
depressed so drastically as to approach $T_R$, in
which case the observed emission quantity  $N_B$ would be much
less than the actual column density $N_{HI}$. As mentioned in the
introduction, ``HI edges'' are observed at $N_B\sim N_l$
and the question arises whether HI could be ``hidden''
outside such edges. In other words, could there be regions
where $N_{HI}$ is only slightly smaller than $N_l$ but
$N_B << N_{HI}$ ? We now give two arguments,
one observational and one theoretical, to show that this is not
the case.

We selected a few objects with low 21-cm surface brightness
($N_B<2\times10^{19}$cm$^{-2}$) observed as part of
outer disks by CS (4 in M33 and 1 in NGC 4631) or as HVC by CTS.
These are good candidates for a phenomenon of $N_{HI}>>N_B$.
In Figure 2
we show the lower limits to $T_S$ calculated from the absence
of 21-cm absorption (we assume a line width of 4 km/s except
for one case in M33 where $T_S > 300$K and we use 8 km/s).
The curves give the $N_{HI}-T_S$ relation (see eq. (4))
for each selected object and the thick part of the curves displays
the allowed values of $N_{HI}$. Since they
are all on the vertical portion of the curve, where $T_S>>T_R$
and $N_{HI} \simeq N_B$, we can exclude that strong subthermal
effects hold for these cases.

For the lowest extragalactic flux and pressure, namely Case A
with $E_c=1.5$ or 0.5 keV and $P_0/k=0.1$ cm$^{-3}$ K,
we have calculated $T_S$.
These small values of flux and pressure should give the
largest difference between $N_B$ and $N_{HI}$.
Figure 3 shows the $N_B-N_{HI}$ relation and
we can see that no drastic depressions in the brightness are
expected for $N_B> 10^{18}$ cm$^{-2}$ (appreciably smaller
than $N_l$).
Therefore this proves again that the surface brightness is
still a good indicator of the neutral hydrogen column density
where HI edges occur. Our conclusion that HI edges
are not due to subthermal effects holds even more strongly
if the background spectrum extends down to lower energies.
Table 1 gives $N_{HI}/N_B$, $T_K$ and $T_S$
for Case A with $E_c=1.5$ and 0.5 keV and for a few low
values of $P_0$.
Notice that the gas is always in the cold phase
and as we increase the pressure $T_S$ approaches $T_K$,
$T_K$ decreases until reaches 12.5 K at very large $P_0$.
This is the lowest possible temperature for our assumed metal
abundance and it is reached when heating is provided
by carbon ionization.
For larger ionizing fluxes (e.g. Case C) we have
larger values of $T_K$ and of $T_S$ with a consequent decrease of
subthermal effects in emission.
Values of $T_S$ in Table 1 are already too large to allow
substantial departures of $N_B$ from $N_{HI}$, but they are
smaller than those
observed by CS and CTS (some of which are shown in Fig. 2).
This implies extra heat sources in outer regions
(Case B) or larger ionizing fluxes than we used for Case A
(case C).

\vglue 0.1 true in

\section
{THE SPIN TEMPERATURE - HEAT RELATION}

\vglue 0.1 true in

The flux assumed in
Case A is unrealistically small and implies
spin temperatures well below those observed. We confine
therefore our attention to
Case B and Case C which
involve stronger energy inputs to the gas than Case A.
In Case B the ionizing flux is fixed, the non radiative
energy input rate $E_{nr}$ varies, and $T_S$ (as well as $T_K$)
are calculated. In Case C, $E_{nr}=0$ and the intensity of
the ionizing flux for $E>0.1$ keV, $I$, varies.
The results we show are useful for interpreting present and
future HI absorption data in terms of heat inputs to the gas from
local sources or, in the absence of these, from a cosmic
background flux. Using the HI data we have available at the moment,
we show that the heat input required from
local sources, $E_{nr}$, is small and compatible with the absence
of star formation in outer regions. If there are no local
sources (Case C) a comparison between the observed and predicted
spin temperature for a given HI column density sets interesting
constraints on the intensity of the cosmic background radiation.

We know that for given $P$ and
$T_K$ a medium with a lower ionization fraction in the cold phase
has a lower cooling rate per atom (Spitzer 1978).
This means that for a given $T_K$ Case B requires a heating rate
per atom (radiative plus non radiative) lower than Case C. However,
because subthermal effects in Case B are stronger than in Case C,
this inequality might not hold for a given $T_S$.
Our results show that even when
the external pressure is low and subthermal effects are strong,
there is relatively little difference between the heat required
in Case B and in Case C for a given $T_S$.
Figure 4 plots $E_{nr}$ versus $T_S$ for Case B in the stable
cold and warm phases (solutions requiring $E_{nr}<0$ have been
discarded).
For Case C, Figure 5 plots the parameter $I$ in eq. (3) versus
$T_S$ for $s=2.4$. Notice that for very small values of $I$
the extragalactic radiation field is unimportant in the
energy equation, $T_S$ is constant and determined by the balance
between the carbon heating and
the line cooling. In Figure 6 we plot $T_K$ versus $T_S$ for all
the cases shown in Figure 4 and Figure 5. Notice that the warm
phase can starts with lower spin temperatures than the final
part of the cold phase.

Figure 4 and 5 show $E_{nr}$ or $I$ versus $T_S$ for three
different values of $N_{HI}$ and two values of the additional
pressure $P_0$. As we increase the parameter $P_0$, and therefore
the density, the heat input ($E_{nr}$ or $I$) required to reach a
certain $T_K$ increases but subthermal effects get smaller and
$T_S$ closer to $T_K$. The result for a given $T_S$ is that
$E_{nr}$ or $I$ increases rapidly with $P_0$ if $T_S$ is low.
For high $T_S$ the heat input required
is almost independent of $P_0$ and, if the density is not
big enough that $T_S$ stll differs from $T_K$, it can increase
with $P_0$ (see for example Fig.5 for $T_S\simeq 1000$K).

For Case C the existence of the cold and/or warm phase of HI
depends mainly on the ratio of $P$ (from eq. (7)) to the intensity
$I$: for $P/I < 3k$ there is no cold phase at all, for
slightly larger values both phases coexist, and for $P/I >
10k$ there is no warm phase. If the contribution to $I$ from
known quasars alone gives $I\sim 5$, the outermost regions
of the HI disk, with column densities $N_{HI} \sim 3
\times 10^{19}$ cm$^{-2}$, are likely to be all in the warm phase
when $z_{1/2} \sim 1$ kpc. This should be
true even in the absence of $E_{nr}$ or other additional
ionizing fluxes (a warm phase
with smaller $z_{1/2}$ requires stronger background fluxes).
This value of $z_{1/2}$ is what we estimate
from Merrifield (1992) for the outermost disk of the Milky Way.
Therefore even if the outermost part of galactic HI disks are warm
there is no strong evidence for the heat
input in outer regions to be much larger than that given
by Madau's flux. Figures 4,5 and 6 show
a curious feature for Case B with low heat input
($E_{nr} \simlt 10^{-15}$ eV s$^{-1}$) and for Case C with
$I\simlt 10$: the hydrogen is all in the warm phase
with $T_K\sim 10^4$K but the spin temperature is surprisingly low,
$T_S \simlt 500$K. Conversely, if observations should give
$T_S \simgt 1000$K much larger ionizing
fluxes and/or larger $E_{nr}$ would be required.

We illustrate the use of Figures 4, 5, and 6 by applying them to
the analysis of the
observational data for the strongest background source behind the
outer disk of M33 (the curve farthest to the right in Fig. 3).
Here we have $N_{HI} \simeq N_B \simeq 2 \times 10^{19}$
cm$^{-2}$ along the line of sight, and the absence of absorption
gives $T_S > 250$ K
(Corbelli $\&$ Schneider 1990). The HI surface density
perpendicular to the plane will be slightly smaller
than the observed value due to the inclination of the
outer disk respect to our line of sight (which is less
than that of the inner disk but not completely
negligible, see Corbelli {\it et al.} 1989).
Fig. 4 $(b)$ and 5 $(b)$ show that we have possible
solutions for both Case B and C with small heat input
if $P_0$ is reasonably small ($P_0/k \simlt 10$, say).
The minimum heat compatible with $T_S \sim 250$ K for
small pressures is similar for Case B and C and gives
solutions in the warm phase with $E_{nr}\simeq
5\times 10^{-16}$ eV sec$^{-1}$ for Case B and $I \simeq
6$ for Case C with $s=2.4$ and $E_{c}=0.1$ keV.
This value of $I$ can be achieved by quasars alone
(Madau 1992) and this value of $E_{nr}$ can be supplied by
a mild outer galactic fountain (Corbelli $\&$ Salpeter
1988). For instance, one
requires only 1$\%$ of the total galactic supernova energy output
rate to flow with energy of $\sim 10^5$ eV cm$^{-2}$ s$^{-1}$
over a disk of 30 kpc radius. If half of this energy is used for
heating the disk gas layer below, then $T_S\sim 500$ K for
$N_{HI}\approx 2$ or $3\times 10^{19}$ cm$^{-2}$. Slightly smaller
spin temperatures are predicted for a gas with lower $N_{HI}$.
To summarize: the present limit of $T_S>250$K is still compatible
with present estimates of the quasar flux and/or a very mild outer
galactic fountain, which are sufficient to keep a gas with
$N_{HI}\simlt 3\times 10^{19}$ cm$^{-2}$ in the
warm phase. The
spin temperatures are surprisingly low, compared to $T_K\sim 10^4$K
(unless the energy input is extremely large).

Another reliable absorption measurement of 21-cm
line outside the optical disk is in the spectrum of the quasar
3C232, at the
velocity of the spiral galaxy NGC3067, close to the quasar
on the plane of the sky.
The width of this line (Rubin {\it et al.} 1982)
implies that the absorbing material is HI in its cold phase
with $T_S \le 300$ K and with $N_{HI}^{cold} \le 5.8\times
10^{19}$cm$^{-2}$ along the line of sight as derived from the
strength of the absorption line. Emission measurements (Carilli
$\&$ van Gorkom 1992) give a total observed HI column density
of ($8\pm4)\times 10^{19}$ cm$^{-2}$; assuming a modest inclination
correction factor, it should be $N_{HI}\sim5\times10^{19}$cm$^{-2}$,
so that Figures 4$(c)$ and 5$(c)$ apply.
We can be in the narrow range of conditions where the warm and
the cold HI phase coexist. If $z_{1/2}$ is of order 1 kpc,
as observed in the Milky
Way, then $P_0/k\sim 30$; in the vicinity of this pressure
Fig. 5$(c)$ gives a flux with $I\sim 3(P_0/10k)^{0.5}$ and Figure
4$(c)$ gives for Case B a heating rate $E_{nr}\sim 5\times 10^{-16}
(P_0/10k)^{0.3}$ eV s$^{-1}$. These are only order of magnitude
relation, but they are compatible with the estimates given above for
M33 and with likely quasar fluxes. Figure 4$(c)$ and 5$(c)$ also make
predictions about the warm phase: $T_S$ should be $\sim 300 - 500$ K
for Case B and only slightly larger for Case C. More than half of the
HI should be in this phase and one might be able to observe a second
absorption component with the peculiar combination of a low spin
temperature and a large width, $\omega \sim$ 20 km s$^{-1}$,
corresponding to the larger $T_K$.

\vglue 0.1 true in

\section
{DARK MATTER DECAY AND HI TEMPERATURES IN OUTER REGIONS}

\vglue 0.1 true in

The dark matter decay theory recently developed by Sciama
(Sciama 1990, 1991), predicts a universe populated by
heavy neutrinos which decay radiatively producing photons
with energy $E_{\gamma}=13.6+\epsilon$
with $\epsilon \sim 1$ eV and a production rate proportional to
the density of dark matter. If the local density of the gas is
sufficiently high, these photons are produced and absorbed
locally and they are the most important photons for the
ionization of HI. In the outermost regions of galactic
disks the local conditions are very different and the monochromatic
photons, both extragalactic and local, can cause a sharp
HI edge, even in the absence of quasar photons. This is
discussed in Paper II. Here we again consider only the mostly
neutral portion of the galactic disk, just inside the edge.
We assume first a
negligible background flux from quasars and no heat source
besides the local neutrino decay photons.  Due to the larger
number of photons with energy close to 13.6 eV, the Lyman-$\alpha$
pumping will be very effective to bring $T_S$ close to $T_K$.
As source function
for these photons we use the expression given by
Sciama (1990) for our Galaxy:

$$\phi
 = {5\times 10^{-16}\over 3\times(1+R/R_0)^2}
{\hbox {cm}}^{-3} {\hbox {s}}^{-1}\eqno(12)$$

\noindent
The equation of state relates
density and pressure; the ionization recombination equilibrium
and the energy equation in the absence of any external flux are:

$${P x_H \over T_K (1+x_H+h_{He})} = \sqrt{\phi \over
\alpha_2}\eqno(13)$$

$$ {P \over T_K (1+x_H+h_{He})}\Big\lbrack \sqrt{\phi \over \alpha_2}
f_1 + (1-x_H){P \over T_K (1+x_H+h_{He})}f_2 \Big\rbrack =
\phi \epsilon \eqno (14)$$

\noindent
Combining these two equations we have a direct relation
between $P$ and $T_K$, which depends on the square root of
the source function $\phi$, and on $\epsilon$
through $x_H$. For $R/R_{0}=3$ Figure 7$(a)$
shows this relation and Figure 7$(b)$ the
corresponding fractional ionization $x_H$.
Notice that the gas with low HI
column density can survive the external flux in the neutral
state only if its pressure is higher than 4 cm$^{-3}$ K;
there is also a range of pressure which allows two different
equilibrium temperatures.
Absorbing the external flux, $F_{ext}$ due to
neutrino decay in the halo and in the rest of the Universe,
requires a column density larger than $N_{min}\equiv
2 F_{ext} T_K/(P \alpha_2)$, which is plotted as the thick
curve in Fig. 7$(c)$. The thin curve in the same Figure shows
the maximum column density compatible with a given $T_K$ (obtained
by setting $P_0=0$).
Therefore in the absence of any other heat source $T_K \simlt 500$
K and the region between the two curves indicates all the possible
$N_{HI}-T_K$ pairs.

If there are extra heat sources in the medium, $T_K$ can be higher
than the values shown in Figure 7 and the energy equation will be
eq. (11) with the
helium terms neglected and the heat from a quasar
background replaced by the heat from the local monochromatic
flux. The non radiative heat input which is needed is:

$$ E_{nr}={P(1-x_H)\over T_K(1+x_H)} f_1+\Bigl\lbrack f_2+x_H f_3 +
(1-x_H)f_4 \Bigr\rbrack {\phi_{\nu} \over \alpha_2} - \epsilon \phi
{T_K (1+x_H)\over P} \eqno(15)$$

\noindent
where $x_H=\sqrt{\phi_{\nu}/\alpha_2}/(P/T_K-\sqrt{\phi_{\nu}/
\alpha_2})$.
$E_{nr}$ depends on $N_{HI}$ only through the pressure term
and therefore we can plot $E_{nr}$ versus $T_K$ for several
values of $P$ without using $P_0$ and $N_{HI}$ explicitly.
Figure 8$(a)$ shows the $E_{nr}-T_K$ relation for $P/k=10,
100,1000$ cm$^{-3}$ K. The extent of the curve is limited by
the conditions
$E_{nr}>0$, $0<x_H<1$ and $N_{min}<N_{HI}<N_{max}$ ($N_{min}$ and
$N_{max}$ being defined as in Figure 7$(c)$).
For the same values of $P$, Figure 8$(b)$ shows $N_{min}$ (thick
curves) and $N_{max}$ (thin curves) as functions of $T_K$.
Suppose we have gas with $N_{HI}\simeq 2\times 10^{19}$ cm$^{-2}$
and $T_S>500$K; then the minimum required extra heat will be $E_{nr}
\simeq 2 \times 10^{-15}$ eV s$^{-1}$ which is comparable to
the $E_{nr}$ required for the same column density by Case B,
where there are no local monochromatic photons but a background
at higher energies. For this case the minimum value of
$P/k$ needed for HI survival against the neutrino's external flux
is $\simeq$ 10 cm$^{-3}$ K. Higher column densities
or higher pressures require slightly higher values of
$E_{nr}$. These $E_{nr}$ values are in reality upper limits
since even a modest quasar background, which we have neglected,
will contribute somewhat to the heat.

We return now to the two examples of spin temperature measurements.
For $N_{HI}\approx 2\times 10^{19}$ cm$^{-2}$ as in M33, we only
have the observational limit $T_S>250$K. Figure 7$(c)$ shows that
this is compatible with neutrino decay theory even in the absence
of extra heat sources. However for a surface density
$N_{HI}\approx 5\times 10^{19}$ cm$^{-2}$ (as it might be in the
case of NGC3067/3C232) Figure 7$(c)$ gives $T_S<30$K ($T_K\simeq
T_S$)
whereas the observations showed a higher spin temperature
and probably some warm HI. Non-ionizing heat sources are
therefore required, but
$E_{nr}$ need not be larger than for Case B.

\vglue 0.1 true in

\section
{SUMMARY AND DISCUSSION}

\vglue 0.1 true in

In this paper we have discussed HI emission and absorption studies at
21-cm wavelengths which can be used to constrain the spectrum of
the extragalactic
ionizing radiation below 1 keV, and more generally the rate of heat
input outside the optical disk of galaxies. Taking into account
self-gravity and considering only unreasonably small lower
limit to the extragalactic flux
(Case A) we have shown that the number of ionizations and collisions
are sufficient to bring the spin temperature well above
the background temperature $T_R$ even in
the absence of strong compression (due to coronal gas
or dark matter).
This means that the actual column density, $N_{HI}$, does not differ
appreciably from
the value $N_B$, inferred from the 21-cm brightness temperature at the
periphery of galaxies.
HI edges observed in outer disks around
$N_{l}\simeq 2\times 10^{19}$ cm$^{-2}$ (Corbelli {\it et al.}
1989; van Gorkom 1991), where $N_B$ drops very rapidly below
the minimum detectable value, or the small quantity of HVC gas
with $N_{HI}\simlt 5\times 10^{18}$ cm$^{-2}$,
correspond then to real cut-offs in the neutral phase of the
hydrogen distribution and not to a sudden
decrease of the HI spin temperature.
These edges are most likely due to ionization by the
background radiation and occur when the total gas column density drops
below a critical value. This phenomenon will be analyzed in Paper II.

We have carried out model calculations for both
the gas kinetic temperature, $T_K$, and the HI spin temperature,
$T_S$, for various values of gas density and
extragalactic background fluxes. We consider a fairly
narrow range of hydrogen column densities appropriate to
outermost HI disks of spiral galaxies, $N_{HI}\sim$ (0.5 to 5)
$\times 10^{19}$ cm$^{-2}$, which are most sensitive to photon
energies $\sim 0.1$ keV. We consider fluxes
from below to above the current range of estimates of
quasar backgrounds at zero redshifts. As in the classical
``two-phase model'' we find for these backgrounds
$T_K$ $\sim$100K for the
cold phase and  $T_K$ $\sim10^4$K for the warm phase.
Deviations from thermal equilibrium are still strong and
$T_S<<T_K$ although $T_S>>3$K always. We have carried out
calculations both with and
without additional non-ionizing sources, $E_{nr}$ (to model
mild galactic fountains, hydromagnetic waves, etc., traveling
from an inner disk to an outer disk).
The relationships between $E_{nr}$, $N_{HI}$, $T_S$, $T_K$ and
the intensity $I$ of the background flux are given in Figures
4, 5, and 6.

The models depend not only on the intensity $I$ (for $E_{nr}=0$),
but also on an additional term $P_0$, with which we estimate the
effective compression of the gas (dark matter compression
versus expansion due to some
internal non thermal pressure). If we
assume that outermost HI disks with $N_{HI} \sim $(2 to 5)$\times
10^{19}$ cm$^{-2}$ have a vertical scale height $z_{1/2}\sim
1$ kpc, as found  for our Galaxy (Merrifield 1992), we
obtain $P_0 \sim$ (10 to 30) cm$^{-3}$ K. Our models predict
that the HI gas in low column density regions should be
in the warm phase but the gas in regions with column density
of order 5$\times 10^{19}$ (which have a slightly higher volume
density) should be a mixture of cold and warm phases. We analyzed
data from an outer
region of M33 with $N_{HI}\sim 2\times 10^{19}$ cm$^{-2}$ where
only an upper limit to the absorption was found (Corbelli $\&$
Schneider 1990) and one region in NGC3067 with $N_{HI}\sim
5\times10^{19}$ where some absorption was detected (Carilli $\&$
van Gorkom 1992). The non detection of 21-cm absorption lines
through the outer disk of M33, which is a relatively
undisturbed HI disk, points out that a significant fraction of the HI
gas is warm despite the absence of star-forming regions (in
the inner, star-forming disk the fraction of cool HI for M33 is only
slightly smaller than for the Milky Way; Dickey $\&$ Brinks 1993).
This requires a cosmic background stronger
than given by the extrapolation of the hard X-ray
spectrum down to 0.1 keV.
The absorption/emission study for NGC3067 suggests that both the
warm and the cold HI phases are present at the slightly larger
column density there.

We found that both observations are compatible
with an intensity $I$ of extragalactic photons at energies $\sim$
0.1 keV roughly as estimated by Madau (1992) for the present day
quasar background.
If this extragalactic ionizing flux should turn out
to be smaller, one can still explain the observed temperatures
by adding a modest amount of non ionizing heat input $E_{nr}$.
Similar heat inputs are required if one replaces the external
ionizing flux by monochromatic UV photons produced locally
by neutrino decay (Sciama 1990).
Only if future observations will
show that HI spin temperatures at the periphery of a galaxy are as
large as 1000 K, would strong local heat sources be required in
addition to a QSO ionizing background.

The warm HI phase of the ISM is a dominant component
in outermost regions and we can make the following prediction:
besides possible narrow HI absorption
lines from any cold phase, the warm HI should have a spin temperature
still very different from the kinetic one. Values of
$T_S \sim 500$K should be common  unless there are
heat inputs, either from a background or from local sources,
much stronger than what is now believed. This warm phase
should produce an absorption line component
with quite large width ($\omega\sim$ 20 km s$^{-1}$), which is
difficult to detect but would provide an interesting diagnostic
because of the $T_S-\omega$ mismatch.

\heading {ACKNOWLEDGMENTS}

We are grateful to A. Ferrara, C. Giovanardi,
R. Reynolds, M. Roberts, S. Schneider,
D. Sciama, J. H. van Gorkom and to the referee
for useful comments to the original manuscript. This work
was supported in part by NSF grants AST 91-19475, INT 89-13558,
and by the Agenzia Spaziale Italiana.

\vfill
\eject

Table 1

\vfill
\eject

\def\refindent{\advance\leftskip by 24pt \parindent=-24pt}

\heading {REFERENCES}

\vglue 0.1 in

\refindent
Bonilha, J. R. M., Ferch, R., Salpeter, E. E., $\&$ Slater, G.
1979, ApJ, 233, 649.\par

Bowyer, S. 1991, Ann. Rev. Astr. Ap., 29, 59.\par

Brinks, E. 1990, in The Interstellar Medium in Galaxies,
ed. H. A. Thronson
$\&$ J. M. Shull (Dordrecht:Kluwer), p. 39.\par

Brown, R. L. 1971, ApJ, 164, 387.\par

Carilli, C. L., van Gorkom, J. H., $\&$ Stoke, J. T. 1989, Nature, 338,
134.\par

Carilli, C. L., and van Gorkom, J. H. 1992, ApJ, 399, 373.\par

Charlton, J. C., Salpeter, E. E., $\&$ Hogan, C. J. 1993, ApJ, 402,
493.\par

Colgan, S. W. J., Salpeter, E. E., $\&$ Terzian, Y. 1990, ApJ, 351,
503.\par

Corbelli, E., $\&$ Salpeter, E. E. 1988, ApJ, 326, 551.\par

Corbelli, E., $\&$ Salpeter, E. E. 1993, ApJ, submitted (Paper II).\par

Corbelli, E., Schneider, S. E., $\&$ Salpeter, E. E. 1989, AJ, 97,
390.\par

Corbelli, E., $\&$ Schneider, S. E.  1990, ApJ, 356, 14.\par

Deguchi, S. $\&$ Watson, W. D. 1985, ApJ, 290, 578.\par

Dickey, J. M. $\&$ Brinks, E. 1993, ApJ, 405, 153.\par

Ferriere, K. M., Zweibel, E. G., $\&$ Shull, J.M. 1988, ApJ, 332,
984.\par

Field, G. B. 1958, Proc. Inst. of Radio Engineers, 46, 240.\par

Giovanelli, R. 1980, AJ, 85, 1155.\par

Giovanelli, R., $\&$ Haynes, M. P. 1988, in Galactic and Extragalactic
Radio
Astronomy, ed. G. L. Verschuur $\&$ K. I. Kellerman
(New York:
Springer-Verlag), p. 522.\par

Giovanelli, R., $\&$ Haynes, M. P. 1991, ApJ, L5.\par

Hasinger, G., Schmidt, M., $\&$ Trumper, J. 1991, AA, 246, L2.\par

Hartwick, F. D. A., $\&$ Shade, D. 1990, ARA$\&$A, 28, 437.\par

Heiles, C. 1987, ApJ, 315, 555.\par

Hoffman, G. L., Lu, N. Y., Salpeter, E. E., Farhat, B., Lamphier,
$\&$ Roos T. 1993, AJ, submitted.\par

Iba\~nez, S. M. H., $\&$ di Sigalotti, L. 1984, ApJ, 285, 784.\par

Knapp, G. R. 1990, in The Interstellar Medium in Galaxies,
ed. H. A. Thronson
and J. M. Shull (Dordrecht:Kluwer), p. 3.\par

Kulkarni, S. R., $\&$ Heiles, C. 1988, in Galactic and Extragalactic
Radio Astronomy,
ed. G. L. Verschuur and K. I. Kellerman (New York:Springer-Verlag),
p. 95.\par

MacLow, M. M., $\&$ McCray, R. 1988, ApJ, 324, 776.\par

Madau, P. 1992, ApJ, 389, L1.\par

Martin, C., $\&$ Bowyer, S. 1990, ApJ 350, 242.\par

McCammon, D., $\&$ Sanders, W. T. 1990, ARA$\&$, 28, 657.\par

McKee, C.F., $\&$ Ostriker, J. P. 1977, ApJ 218, 148.\par

Melott, A. L., McKay, D. W., $\&$ Ralston, J. P. 1988, ApJ, 324, L43.\par

Merrifield, M. R., 1992, AJ, 103, 1552.\par

Reynolds, R. J. 1990, ApJ, 348, 153.\par

Rubin, V. C., Thonnard, N., $\&$ Ford, K. W. 1982, AJ 87, 477.\par

Sargent, W. L. W., Young, P. J., Boksenberg, A., Carswell, R. F.,
$\&$ Whelan,
J. A. J. 1979, ApJ, 230, 49.\par

Schneider, S. E., Skrutskie, M. F., Hacking, P. B., Young, J. S.,
Dickman, R. L.Claussen, M. J., Salpeter, E. E., Houck, J. R.,
Terzian, Y., Lewis, B. M.
$\&$ Shure, M. A. 1989, AJ, 97, 666.\par

Schneider, S. E., $\&$ Corbelli, E. 1993, ApJ, 414, in press.\par

Schwartz, D. A. 1978, {\it in X-ray Astronomy} ed. W. A. Baity and
L. E. Peterson
(Oxford: Pergamon) p. 453.\par

Schwartz, D. A., $\&$ Tucker, W. H., 1988, ApJ, 332, 157.\par

Sciama, D. W. 1990, ApJ, 364, 549.\par

Sciama, D. W. 1991, A$\&$A, 245, 243.\par

Shanks, T., Georgantopoulos, I., Stewart, G. C., Pounds, K. A.,
Boyle, B. J.
$\&$ Griffiths, R.E. 1991, Nature, 353, 315.\par

Shull, J. M., $\&$ Van Steenberg, M. E. 1985, ApJ, 298, 268.
508.\par

Smith, F. J. 1966, Planet. Space Sci., 14, L71.\par

Spitzer, L. 1942, ApJ, 95, 329.\par

Spitzer, L. 1978, Physical Processes in The Interstellar
Medium, (Wiley: New York).\par

van Gorkom, J. H. 1991, ASP Conference series n.16, (proceedings
3rd Haystack Observ. Conference on Atoms, Ions and Molecules,
ed. A. D. Haschick and P. T. P. Ho), p. 1.\par

Wakker, B. P., Vijfschaft, B., $\&$ Schwarz, U. J. 1991, A$\&$A,
249,  L5.\par

Wu, X., Hamilton, T. T., Helfand, D. J., $\&$ Wang, Q. 1991,
ApJ, 379, 564.\par

Wu, X., $\&$ Anderson, S. F. 1992, AJ, 103, 1.\par

\vfill
\eject

\heading
{FIGURE CAPTIONS}

\vglue 0.1 in

{\bf Figure 1}. Some of the X-ray spectra used. At energies
above 1.5 keV all the spectra have $s=1.4$
$I=7.7$, and
the thick line shows this spectrum extended down to 0.1 keV as it
has been used for Case B. The same spectrum has been used for Case A
but with lower cut off energies at 1.5 or 0.5 keV.
All the other lines are examples of
spectra used for Case C and are labelled according to the corresponding
values of $I$ and $s$ (see eq. (3)).

{\bf Figure 2}. The $T_S - N_{HI}$ relation
for all the positive detections of HI with $N_{B}\le 2 \times
10^{19}$ cm$^{-2}$ in
CS (circles) and $N_{B}\le 10^{19}$ cm$^{-2}$ in CST (triangles).
The thick
parts of the curves show the possible pairings
of $N_{HI}$ and $T_S$ for each data point according to the upper
limits obtained from 21-cm absorption data.

{\bf Figure 3}. Subthermal effects in emission: the column density
derived directly from the brightness temperature, $N_B$, versus
the real $N_{HI}$ of the gas.  We have
used $P_0=0.1$ cm$^{-3}$K, and a low
external flux, namely Case A with $E_c=1.5$ keV
(dotted line) or $E_c=0.5$ keV (dashed line). The
solid line is the line of unity slope in the $N_B$-$N_{HI}$ plane.

{\bf Figure 4}. Heat per atom required for equilibrium
in units of eV s$^{-1}$ for Case B ($E_{c}=0.1$ keV, $s=1.4$,
$I=7.7$). It is shown as function of $T_S$
in the stable cold and warm phases for two values of $P_0$.
The filled triangles indicate the
beginning of the two phases when $P_0/k$=1 cm$^{-3}$ K (solid
lines) and the open triangles when $P_0/k=100$ cm$^{-3}$ K
(dashed lines). $N_{HI}$ = $5\times 10^{18}$ cm$^{-2}$ in $(a)$,
$2\times 10^{19}$ cm$^{-2}$ in ($b$), and
$5\times 10^{19}$ cm$^{-2}$ in ($c$).

{\bf Figure 5}.
The parameter $I$ of the spectrum in
eq. (3) as a function of $T_S$ for
Case C when $s=2.4$. The filled triangles indicate the
beginning of the cold and warm phase when $P_0/k$=1 cm$^{-3}$ K
(solid lines) and the open triangles when $P_0/k=100$ cm$^{-3}$ K
(dashed lines). $N_{HI}$ is $5\times 10^{18}$ cm$^{-2}$ in $(a)$ is
$2\times 10^{19}$ cm$^{-2}$ in ($b$) and is
$5\times 10^{19}$ cm$^{-2}$ in ($c$)

{\bf Figure 6}.
The kinetic temperature $T_K$-$T_S$ relation for
all cases shown in Fig. 4 and in Fig. 5. $(a),(b),(c)$
are for Case B with $N_{HI}$ = $5\times 10^{18}$,
$2\times 10^{19}$, and  $5\times 10^{19}$ cm$^{-2}$
respectively. For the same values of  $N_{HI}$
$(e),(f),(d)$ show the  $T_K$-$T_S$ relation for
Case C with $s=2.4$ and $E_c=0.1$ keV. Lower curves
in the panels are for the cold phases while curves
at the top of the panels are for the warm phases.

{\bf Figure 7}. In $(a)$ and $(b)$ we show the pressure and the
hydrogen fractional ionization for a gas pervasive of monochromatic
photons of 14.6 eV from decaying neutrinos and no other heat sources.
The thin and thick curves in $(c)$ show the maximum and
minimum HI column density, respectively, which are
compatible with the corresponding value of $T_K$ (see text for
details).

{\bf Figure 8}. In $(a)$ the extra heat per atom required by a gas
with $P=10$ (continuous line), $P=100$ (small dashed line) and
$P=1000$ (large dashed line) cm$^{-3}$ K, is shown as function
of the kinetic temperature. In $(b)$ thin and thick curves
are the maximum and minimum HI column density respectively, which are
compatible with a given value of $T_K$ (see text for details).

\end